\documentclass[5p,number]{elsarticle}
\usepackage{amsmath}
\usepackage{amssymb}
\usepackage{graphicx}
\usepackage{bm}
\usepackage{natbib}

\journal{Physics Letters B}

\begin{document}

\begin{frontmatter}

\title{First Measurement of the Neutral Current Excitation of the $\Delta$ Resonance on a Proton Target}

%Full G0 author list
\author[uzag]{D.~Androi\'c} 
\author[wm]{D.~S.~Armstrong} 
\author[ors]{J.~Arvieux \fnref{fn1}} 
\author[wm]{S.~L.~Bailey} 
\author[uiuc]{D.~H.~Beck} 
\author[umd]{E.~J.~Beise} 
\author[jlab]{J.~Benesch} 
\author[umd,cmu]{F.~Benmokhtar} 
\author[ors]{L.~Bimbot} 
\author[uman]{J.~Birchall} 
\author[jlab]{P.~Bosted} 
\author[umd]{H.~Breuer}
\author[wm]{C.~L.~Capuano \corref{cor1}} \ead{ccapuano@jlab.org}
\author[jlab]{Y.-C.~Chao}
\author[uman]{A.~Coppens} 
\author[triumf]{C.~A.~Davis}
\author[umd]{C.~Ellis} 
\author[nmsu]{G.~Flores}
\author[cmu]{G.~Franklin} 
\author[lpsc]{C.~Furget} 
\author[jlab]{D.~Gaskell}
\author[jlab]{J.~Grames} 
\author[uman]{M.~T.~W.~Gericke} 
\author[lpsc]{G.~Guillard} 
\author[jlab]{J.~Hansknecht} 
\author[jlab]{T.~Horn} 
\author[jlab]{M.~K.~Jones} 
\author[ohio]{P.~M.~King}
\author[uky]{W.~Korsch}
\author[lpsc]{S.~Kox}
\author[uman]{L.~Lee}
\author[caltech]{J.~Liu}
\author[jlab]{A.~Lung}
\author[vt]{J.~Mammei} 
\author[uwin]{J.~W.~Martin}
\author[caltech]{R.~D.~McKeown}
\author[gwu]{A.~Micherdzinska} 
\author[jsi]{M.~Mihovilovic}
\author[ypi]{H.~Mkrtchyan}
\author[uiuc]{M.~Muether} 
\author[uman]{S.~A.~Page} 
\author[nmsu]{V.~Papavassiliou}
\author[nmsu]{S.~F.~Pate}
\author[wm]{S.~K.~Phillips}
\author[lpsc]{P. Pillot} 
\author[vt]{M.~L.~Pitt} 
\author[jlab]{M.~Poelker} 
\author[cmu]{B.~Quinn} 
\author[uman,triumf]{W.~D.~Ramsay}
\author[lpsc]{J.-S.~Real}
\author[ohio]{J.~Roche}
\author[umd]{P.~Roos}
\author[nmsu]{J.~Schaub}
\author[uzag]{T.~Seva}
\author[latech]{N.~Simicevic} 
\author[jlab]{G.~R.~Smith} 
\author[hen]{D.~T.~Spayde}
\author[jlab]{M.~Stutzman} 
\author[vt,jlab]{R.~Suleiman} 
\author[ypi]{V.~Tadevosyan}
\author[uman]{W.~T.~H.~van~Oers} 
\author[lpsc]{M.~Versteegen} 
\author[lpsc]{E.~Voutier}
\author[jlab]{W.~Vulcan} 
\author[latech]{S.~P.~Wells} 
\author[uiuc]{S.~E.~Williamson}
\author[jlab]{S.~A.~Wood}

%Institution list
\address[uzag]{Department of Physics, University of Zagreb, Zagreb HR-41001 Croatia} 
\address[wm]{Department of Physics, College of William and Mary, Williamsburg, VA 23187 USA} 
\address[ors]{Institut de Physique Nucl\'eaire d'Orsay, Universit\'e Paris-Sud, F-91406 Orsay Cedex FRANCE}
\address[uiuc]{Loomis Laboratory of Physics, University of Illinois, Urbana, IL 61801 USA}
\address[umd]{Department of Physics, University of Maryland, College Park, MD 20742 USA}
\address[jlab]{Thomas Jefferson National Accelerator Facility, Newport News, VA 23606 USA}
\address[cmu]{Department of Physics, Carnegie Mellon University, Pittsburgh, PA 15213 USA}
\address[uman]{Department of Physics, University of Manitoba, Winnipeg, MB R3T 2N2 CANADA}
\address[triumf]{TRIUMF, Vancouver, BC V6T 2A3 CANADA}
\address[nmsu]{Department of Physics, New Mexico State University, Las Cruces, NM 88003 USA}
\address[lpsc]{LPSC, Universit\'e Joseph Fourier Grenoble 1, CNRS/IN2P3, Institut Polytechnique de Grenoble, Grenoble, FRANCE}
\address[ohio]{Department of Physics and Astronomy, Ohio University, Athens, OH 45701 USA}
\address[uky]{Department of Physics and Astronomy, University of Kentucky, Lexington, KY 40506 USA}
\address[caltech]{Kellogg Radiation Laboratory, California Institute of Technology,  Pasadena, CA 91125 USA}
\address[vt]{Department of Physics, Virginia Tech, Blacksburg, VA 24061 USA}
\address[uwin]{Department of Physics, University of Winnipeg, Winnipeg, MB R3B 2E9 CANADA}
\address[gwu]{Department of Physics, The George Washington University, Washington, DC 20052 USA}
\address[jsi]{Jo\^zef Stefan Institute, 1000 Ljubljana, SLOVENIA}
\address[ypi]{Yerevan Physics Institute, Yerevan 375036 ARMENIA}
\address[latech]{Department of Physics, Louisiana Tech University,  Ruston, LA 71272 USA}
\address[hen]{Department of Physics, Hendrix College, Conway, AR 72032 USA}

\cortext[cor1]{Corresponding author}

\fntext[fn1]{Deceased}

\date{\today}

\begin{abstract}
The parity-violating asymmetry arising from inelastic electron-nucleon scattering at backward angle ($\sim95^{\circ}$) near the $\Delta(1232)$ resonance has been measured using a hydrogen target. From this asymmetry, we extracted the axial transition form factor $G^A_{N\Delta}$, a function of the axial Adler form factors $C^A_i$. Though $G^A_{N\Delta}$ has been previously studied using charged current reactions, this is the first measurement of the weak neutral current excitation of the $\Delta$ using a proton target. For $Q^2$ = 0.34 (GeV/c)$^2$ and $W$ = 1.18 GeV, the asymmetry was measured to be $-33.4 \pm (5.3)_{stat} \pm (5.1)_{sys}$ ppm. The value of $G^A_{N\Delta}$ determined from the hydrogen asymmetry was $-0.05 \pm (0.35)_{stat} \pm (0.34)_{sys} \pm (0.06)_{theory}$. These findings agree within errors with theoretical predictions for both the total asymmetry and the form factor. In addition to the hydrogen measurement, the asymmetry was measured at the same kinematics using a deuterium target. The asymmetry for deuterium was determined to be $-43.6 \pm (14.6)_{stat} \pm (6.2)_{sys}$ ppm.
\end{abstract}


\end{frontmatter}

\section{Introduction}

A measurement of the parity-violating (PV) asymmetry in inelastic electron-nucleon scattering near the $\Delta$ resonance has been performed as part of the $G^0$ experiment \cite{PACProposal}. The study of the PV asymmetry in $\Delta$ electroproduction using a neutral weak probe was first considered by Cahn and Gilman \cite{CahnGil78} as a test of the Standard Model, due to the strong dependence of the predicted
asymmetry on the weak mixing angle, $\theta_W$. However, it is now clear that uncertainties in the theoretical predictions of the asymmetry arising from hadron structure and radiative corrections prevent a precision Standard Model test using this reaction. Instead, the inelastic asymmetry can be used to study other physics topics, including the axial $N\rightarrow\Delta$ transition. In this paper we report the first measurement of this neutral current excitation of the $\Delta$ with hydrogen and deuterium targets. The axial transition form factor, $G^A_{N\Delta}$, a linear combination of the axial Adler form factors, was determined from the results on hydrogen. 

$\Delta$ production occurs in both the charged current (CC) and neutral current (NC) channels of the weak interaction. In the constituent-quark model for the nucleon, two of the quark spins are aligned while the third is anti-aligned. The $\Delta$ results from interactions that flip the spin of the anti-aligned quark, leading to a total $\Delta$ spin of $J = \frac{3}{2}$. In CC reactions, there is a quark flavor change in addition to the spin flip. Theoretical predictions for the PV asymmetry rely on isospin symmetry and the assumption that the CC form factors will not differ significantly from their NC analogues. A precise measurement of the inelastic PV asymmetry could confirm that these assumptions are valid.

The axial vector response in the $N\rightarrow\Delta$ transition has been studied theoretically using chiral constituent quark models \cite{Barq07}, chiral perturbation theory \cite{Proc08,Geng08}, light cone QCD sum rules \cite{Aliev08}, and lattice gauge theory \cite{AlexPRL07, AlexPRD07, Alex11}. Available data are sparse, however, and the only data from nucleon targets come from CC neutrino experiments using bubble chambers \cite{Kit90,Rad82,Bar79}. This subject is also topical because lack of knowledge of $G^A_{N\Delta}$ is one of the dominant uncertainties in understanding neutral current pion production \cite{Alv07}, an important background in $\nu_e$ appearance neutrino oscillation experiments.

\section{Theoretical Background}
\label{SecTheory}

In the scattering of longitudinally polarized electrons, the interference between $\gamma$ and $Z^0$ exchange amplitudes leads to a parity-violating dependence of the cross section on the electron helicity. One can form an experimental asymmetry 
\begin{equation}
A = \frac{d\sigma_R-d\sigma_L}{d\sigma_R+d\sigma_L}~,
\end{equation}
where $d\sigma_R$ and $d\sigma_L$ are the differential cross sections for scattering of right- and left-handed electrons, respectively. A measurement of 
this asymmetry then allows study of the neutral-current contribution to the reaction. 

The PV asymmetry in electron-proton scattering near the $\Delta$ resonance can be written \cite{PhysRept94}
\begin{align}
  A_{inel}&= \frac{1}{2} A^0 \left [ \Delta^{\pi}_{(1)} + \Delta^{\pi}_{(2)} + 
                                   \Delta^{\pi}_{(3)} \right ]~\\
         \label{AsyDefn}  
         &= A_1 + A_2 + A_3, \nonumber
\end{align}
where the use of the $\pi$ superscript indicates single pion production and $A^0$ is defined as
\begin{equation}
  A^0 = -\frac{G_F Q^2}{2 \pi \alpha \sqrt{2}}~,
  \label{DefnA0}
\end{equation}
where $G_F$ is the Fermi constant, $\alpha$ is the fine structure constant, and $Q^2$ is the negative four-momentum transfer squared ($Q^2 = -q^2$). 
$A^{0}$ evaluates to $-61$ ppm at the experimental kinematics (see Section \ref{SecExp}). The three $\Delta^{\pi}_{(i)}$ terms represent a decomposition of the hadronic response into resonant vector, non-resonant vector, and axial vector contributions, respectively. Here, ``vector'' and ``axial vector'' refer to the Lorentz structure of the hadronic current in the scattering. In the discussion that follows, we present our formalism for $A_{inel}$ which combines formalisms available in the literature to develop our method for the determination of the axial response. The present formalism was primarily derived from the work in Refs \cite{PhysRept94}, \cite{Nath82} and \cite{Zhu01}.

% define dpi1
The resonant vector hadron term, $\Delta_{(1)}^{\pi}$, is the dominant term in the asymmetry and the only one that is not dependent on hadronic structure. It depends only on the weak mixing angle, $\theta_W$. $\Delta_{(1)}^{\pi}$ contains the full contribution of the resonant vector current at the hadronic vertex to the asymmetry and can be written \cite{PhysRept94}
\begin{align}
  \Delta_{(1)}^{\pi} & = g^{e}_A \xi_V^{T=1} \nonumber\\
                   & = 2(1-2\sin^2 \theta_W)~,
  \label{dpi1defn}
\end{align}
where $g^e_A$ is the axial vector coupling of the electron to the $Z$ boson, which is equal to 1 in the Standard Model, and $ \xi_V^{T=1}$ is the isovector hadron coupling to the vector $Z$, which is $2(1-2\sin^2 \theta_W)$ in the Standard Model. Using the world value $\sin^2\theta_W$ = 0.2353 $\pm$ 0.00013 \cite{PDG2010}, $\Delta_{(1)}^{\pi}$ is computed to be 1.06, leading to an $A_1 =-32.2$ ppm contribution to the asymmetry. 

Two different formalisms were applied to the terms $\Delta^{\pi}_{(2)}$ and $\Delta^{\pi}_{(3)}$. The first is a phenomenological approach, outlined by Musolf \emph{et al.} \cite{PhysRept94}.  The second uses the dynamical model of Matsui, Sato, and Lee \cite{TheoryLee}, and is described later.

% define dpi2
In the phenomenological approach, the non-resonant vector term, $\Delta^{\pi}_{(2)}$, is written as a sum of longitudinal, transverse magnetic, and transverse electric multipoles using an isospin decomposition \cite{PhysRept94}. We computed the sum through $l = 2$ using multipoles from MAID2007 \cite{MAID2000}. Further  details on our implementation are available elsewhere \cite{CapuanoThesis}. The theoretical value of $\Delta^{\pi}_{(2)}$ was determined to be 0.018, leading to an asymmetry of $A_2 = -0.55$ ppm. An uncertainty of $\sigma_2^{th}$ = 0.72 ppm is applied to account for estimates made in the model.

% define dpi3
The axial term, $\Delta^{\pi}_{(3)}$, can be written in terms of $F(Q^2,s)$, which contains both axial and electromagnetic form factors,
\begin{align}
  \Delta_{(3)}^{\pi} & \approx g^e_V \xi_A^{T=1}F(Q^2,s) \nonumber\\
                   & \approx 2(1-4\sin^2 \theta_W)F(Q^2,s)~,
  \label{dpi3defn}                   
\end{align}
where $s$ is the Mandelstam invariant and $g^e_V$ and $\xi_A^{T=1}$ have been replaced with their respective Standard Model tree level values of $g^e_V$ = $(-1+4\sin^2 \theta_W)$ and $\xi_A^{T=1}$= $-$2. In Ref. \cite{PhysRept94}, $\Delta^{\pi}_{(3)}$ is defined as the total axial contribution, including both resonant and non-resonant terms. Here, we neglect any non-resonant axial contributions (hence the use of ``$\approx$'' in Equation \ref{dpi3defn}), because theoretical studies \cite{HamDres95} \cite{Mukho98} indicate that these are small. 

The function $F(Q^2,s)$ can be written as a product of two functions of form factors,
\begin{equation}
  F(Q^2,s) = \frac{E + E'}{M} H^{EM}(Q^2,\theta) G^A_{N\Delta}(Q^2)~,
  \label{FQ2sBasic}
\end{equation}
where $H^{EM}(Q^2,\theta)$ and $G^A_{N\Delta}(Q^2)$ are linear combinations of electromagnetic ($C^{\gamma}_i$) and axial ($C^A_i$) form factors, 
respectively, and $M$ is the nucleon mass. The explicit functional form of $G^A_{N\Delta}(Q^2)$ will be presented below,
while $H^{EM}(Q^2,\theta)$ is described in Ref. \cite{CapuanoThesis}. 

% define form factors
In order to calculate a theoretical asymmetry, it is necessary to compute the form factors $C^{\gamma}_i$ and $C^A_i$. One convenient way to express the $Q^2$ dependence of the form factors is through the use of dipole forms. In this notation, referred to as the Adler parameterization \cite{Adler68}\cite{Adler75}, the form factors are expressed as
\begin{align}
  C^{\gamma}_i(Q^2) &= C^{\gamma}_i(0) G^V_D(Q^2)~,  \label{DefnAdlerCV}\\
  C^{A}_i(Q^2) &= C^{A}_i(0) G^A_D(Q^2)\xi^A(Q^2)~,
  \label{DefnAdlerCA}
\end{align}
where the functions $ G^{V,A}_D(Q^2)$ are dipole form factors defined as
\begin{align}
  G^{V,A}_D(Q^2) = \bigg [ 1 + \frac{Q^2}{M^2_{V,A}} \bigg ]^{-2}~.
  \label{InelAsyDipole}
\end{align}
The parameters $M_{V,A}$ are the vector (V) and axial (A) dipole masses, which have been determined from fits to existing data. The current world values for these masses are $M_V$ = 0.84 GeV \cite{VectorMass11} and $M_A$ = 1.03 $\pm$ 0.02 GeV \cite{Bernard02}. 

The function $\xi^A$ is used to give additional structure to the $Q^2$ dependence of the axial response and is written
\begin{equation}
  \xi^A(Q^2) = 1 + \Bigg ( \frac{a'Q^2}{b'+Q^2} \Bigg )~,
  \label{InelAsyXi}
\end{equation}
with the parameters $a'$ and $b'$ determined from a fit to CC neutrino data performed by Schreiner and von Hippel \cite{Sch73}. For the Adler model, $a'$ was found to be $-$1.2 and $b'$ was 2 (GeV/c)$^2$. These results hold only for $Q^2~<$ 0.5 (GeV/c)$^2$, which covers the present experimental kinematics.

The values for $C_i(0)$ are determined from fits to charged current data and are fit-dependent. In this work, the Adler values of these coefficients, as quoted by Nath \cite{Nath82}, were used. They are
\begin{align}
  C^A_3(0) &=  0~,     & C^{\gamma}_3(0) &=  1.85~, \nonumber\\
  C^A_4(0) &= -0.35~,  & C^{\gamma}_4(0) &= -0.89~, 
   \label{CoeffsAtZero}\\
  C^A_5(0) &= 1.20~. \nonumber
\end{align}
Note that not all the $C_i$'s contribute to the asymmetry; $C^{\gamma}_6$ vanishes due to CVC and $C^A_6$ vanishes if we neglect the electron mass. From the coefficients above, we can also neglect $C^A_3$. Further, the photo- and electroproduction data can be fit using the assumption that $C^{\gamma}_5$ = 0 and that $C^{\gamma}_4$ = $-\frac{M}{M+M_{\Delta}}C^{\gamma}_3$ \cite{JonesPetcov80}. Thus, only the $i$ = 3,4 terms of the electromagnetic and the i = 4,5 terms of the axial form factors contribute to the asymmetry. The axial transition form factor expressed in terms of the Adler form factors is then
\begin{align}
 G^A_{N\Delta}(Q^2) = \frac{1}{2}[M^2 - M^2_{\Delta}+Q^2]C^A_4(Q^2) -M^2C^A_5(Q^2)~,
\end{align}
where $M_{\Delta}$ is the mass of the $\Delta(1232)$.

Using this formalism leads to the value $A_3 = -1.8$ ppm for the resonant axial-vector hadron contribution to the asymmetry. We assigned an uncertainty of $\sigma^{th}_3 = 0.65$ ppm to this value to account for the neglect of non-resonant contributions and uncertainties related to the model and fits used to compute $H^{EM}$. Summing the three individual components leads to a total theoretical asymmetry of $A^{th} = -34.6 \pm 1.0$ ppm at the present kinematics.

%EWRC:
Electroweak radiative corrections (EWRC) have been applied to the theoretical asymmetry. We distinguish between `one-quark' and `multi-quark' corrections, adopting the terminology of Zhu \emph{et al.} \cite{Zhu01}. One-quark radiative corrections are those in which the electron interacts only with a single quark in the nucleon, and are calculable with sufficient precision. In contrast, multi-quark corrections for the axial response are largely model dependent.

One-quark EWRC were applied to each $\Delta_{(i)}^{\pi}$ term independently using the corrections reported by the Particle Data Group \cite{PDG2010} for the $\overline{MS}$ renormalization scheme. These corrections have been included in the theoretical asymmetries quoted previously. The one-quark EWRC are small ($< 1.5\%$) for both of the vector hadron terms. For the axial term, however, these EWRC are large, leading to a 58\% reduction in $A_3$.

Zhu \emph{et al.} \cite{Zhu01} have studied the multi-quark axial EWRC and their possible impact on the PV asymmetries. Specifically, they highlighted two effects present at the PV $\gamma N \Delta$ vertex that may contribute to the present measurement. The first effect is an anapole moment analogous to that previously seen in elastic scattering \cite{ZhuAnapole00}. The second, referred to as the Siegert term, is of interest because it could lead to a non-zero asymmetry at $Q^2$ = 0. In a separate analysis, as part of the $G^0$ experiment, we have taken pion photoproduction data on a deuterium target at very low $Q^2$ \cite{G0Pion}. These data allow us to bound the impact of the Siegert term at the present kinematics to $|A_{Siegert}| \leq$ 0.15 ppm. Given the large overall theoretical uncertainty in these effects, multi-quark EWRC are not applied in the present analysis.

% Matsui Sato Lee Formalism:
The second asymmetry formalism is that of Matsui, Sato, and Lee who have developed a dynamical model of pion electroproduction near the $\Delta$ resonance \cite{TheoryLee} and performed a calculation of the inelastic asymmetry at the present kinematics. As in the phenomenological approach, Matsui \emph{et al.} write the asymmetry as a sum of resonant vector, non-resonant vector, and axial vector hadron pieces. In their notation, $A_{inel}$ is given by
\begin{equation}
  A_{inel} = \frac{1}{2} A^0 \bigg[ (2-4\sin^2\theta_W) + \Delta_V +\Delta_A \bigg]~,
\end{equation}
where $(2-4\sin^2\theta_W)$ is identical to $\Delta^{\pi}_{(1)}$ as defined in Equation \ref{dpi1defn} . The two remaining terms, $\Delta_V$ and $\Delta_A$, are equivalent to $\Delta^{\pi}_{(2)}$ and $\Delta^{\pi}_{(3)}$, respectively. However, the formalism used to calculate these terms differs from that which was presented previously. The differences in these two formalisms will be discussed here; more detailed information on the dynamical model is available in Refs \cite{TheoryLee} and \cite{Hemm95}.

For the term $\Delta_V$, Matsui \emph{et al.} derive an expression in terms of structure functions analogous to that used to determine the resonant form of $\Delta^{\pi}_{(3)}$ as presented in Ref \cite{Nath82}. This allows them to determine the contribution of the non-resonant vector hadron reactions to the asymmetry using their dynamical model rather than through the use of phenomenological multipoles.

For the axial term, the definitions of $\Delta_A$ and $\Delta^{\pi}_{(3)}$ in terms of structure functions are the same. Where Matsui \emph{et al.} differ is in their parameterization of the form factors. As before, a dipole form is used for both the vector and the axial vector form factors. However, the additional $Q^2$ parameterization present in the function $\xi^A$ takes on an exponential form rather than that of Equation \ref{InelAsyXi}. Explicitly, $\xi^A$ is given by
\begin{equation}
  \xi^A(Q^2) = (1 + a Q^2)e^{-bQ^2}~,
\end{equation}
where $a$ = 0.154 (GeV/c)$^{-2}$ and $b$ = 0.166 (GeV/c)$^{-2}$ were determined by fits to CC neutrino data. The resulting theoretical values for $A_2$ and $A_3$ computed with the phenomenological approach and the dynamical model each differ by $<1$ ppm. 

\section{Experiment}
\label{SecExp}

The $G^0$ experiment was performed using a beam of longitudinally polarized electrons provided by the accelerator at the Thomas Jefferson National Accelerator Facility. The electrons were scattered from a 20 cm long unpolarized liquid hydrogen or deuterium target. The $G^0$ experiment ran in two phases, collecting data at both forward and backward scattering angles. The inelastic measurement was performed in the backward angle configuration. Since the primary concern of the $G^0$ collaboration was the elastic electron scattering measurement, the experimental apparatus was optimized for elastic kinematics. For the backward angle measurement, the $G^0$ spectrometer consisted of a toroidal magnet, scintillation detectors and aerogel Cherenkov counters. Measurements were taken at beam energies of 687 MeV and 362 MeV, though only the higher of these energies is relevent to the inelastic measurement. Detailed descriptions of each of the components of the $G^0$ experimental apparatus, and their use in both configurations, are given elsewhere \cite{G0NIM}.

Electrons that scattered from the target were bent in a magnetic field and passed through a collimator system before entering the detector system. The collimators defined the experimental acceptance, leading to an effective scattering angle of $\sim95^{\circ}$ for inelastic events and $\sim108^{\circ}$ for elastic events. The detector system was segmented into octants arranged symmetrically around the beamline with each detector octant covering one of the eight gaps between adjacent magnet coils. Each detector octant consisted of two sets of plastic scintillators and an aerogel Cherenkov detector. The scintillators, labeled Focal Plane Detectors (FPDs) and Cryostat Exit Detectors (CEDs), were used for a rough tracking of the scattered electron's path. This led to a two dimensional CED$\cdot$FPD detector space which allowed for a kinematic separation between elastically and inelastically scattered electrons. The Cherenkov detector (CER) was used to distinguish between electrons and pions. A DAQ system consisting of specially-designed logic boards counted coincidences of these detectors, with CED$\cdot$FPD$\cdot$CER coincidences counted as electron events and CED$\cdot$FPD$\cdot\mathrm{\overline{CER}}$  counted as pion events. Thus, electrons and pions were counted side-by-side and their rates recorded separately by scalers. 

\begin{figure}
  \centering  
  \includegraphics[width=\columnwidth]{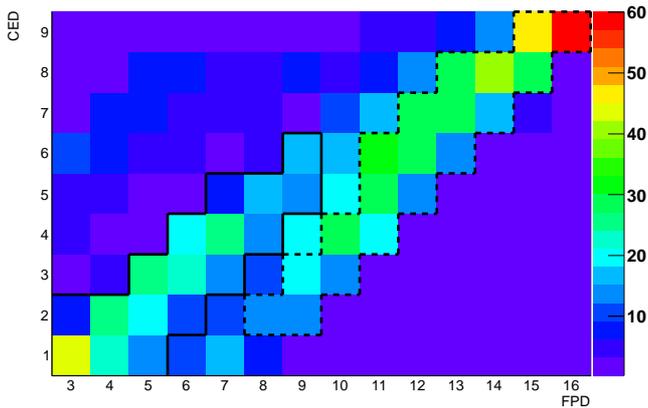}
  \caption{Octant averaged electron yields for the high energy hydrogen run period. On the $y$-axis is CED number and the $x$-axis is FPD number, and each block represents a CED$\cdot$FPD coincidence. The color scale indicates cell yield in Hz/$\mu$A. The collection of cells outlined in with a solid (dashed) line represents the inelastic (elastic) locus. Note that since FPDs 1 and 2 were not used in the backward angle configuration, the FPD numbering begins with 3.}
  \label{CoincMat}
\end{figure}

The helicity of the beam was flipped at a rate of 30 Hz, resulting in a series of 1/30 s segments of common helicity called macropulses (MPSs). The helicity pattern was generated as a collection of four MPSs, referred to as a quartet. The use of quartets, coupled with the fast helicity reversal, cancels linear drifts that can affect the asymmetry. The sequence of the helicity reversal for each quartet was chosen to be either $+--+$ or $-++-$ depending on a randomly generated initial MPS. The coincidence count for each CED$\cdot$FPD pair was recorded for a single MPS, then normalized to the beam current to create an MPS yield. The asymmetry was then computed for each quartet by combining the detector yields for the positive and negative helicity MPSs according to
\begin{equation}
  A_{qrt} = \frac{\sum Y_i^+ - \sum Y_i^- }{\sum Y_i^+ + \sum Y_i^-}~.
  \label{QrtAsyDefn}
\end{equation}
Figure \ref{CoincMat} shows the octant-averaged electron yields for the hydrogen data for each CED$\cdot$FPD coincidence cell. Coincidence cells with common kinematic acceptance were grouped into loci. The inelastic and elastic loci are indicated on the figure. The measured asymmetry for a given locus is taken as the average of the asymmetries in each locus cell, weighted by statistics.

The average kinematics for the inelastic measurement were determined through the use of a GEANT3 simulation \cite{GEANTDoc}. The detector acceptance is defined by the collimators that are part of the spectrometer. For events in the inelastic locus, the range of accepted scattering angles was $85^{\circ} < \theta < 105^{\circ}$ and the average $\theta$ = 95$^{\circ}$. This results in a $Q^2$ range of $0.25 < Q^2 < 0.5$ (GeV/c)$^2$ with an average of $Q^2$ = 0.34 (GeV/c)$^2$. The range of invariant mass, $W$, covers the entire region from the pion threshold to just past the peak of the $\Delta$, with a range of $1.07 < W < 1.26$ GeV. Most of the events were on the low $W$ side of the peak, leading to an average just below the resonance at $W$ = 1.18 GeV. 

\section{Data Analysis}

% Intro Comments
In order to determine the physics asymmetry, several corrections were applied to the raw asymmetry. An overview of the corrections will be provided in this section, while more detailed descriptions of the data analysis related to beam and instrumentation \cite{BckAngleLong}, and the inelastic result \cite{CapuanoThesis} are available elsewhere. A summary of the corrections applied and their contributions to the systematic uncertainty is given in Table \ref{TabCorSum}. 

%RATE CORRECTIONS: 
There were three rate-dependent effects that needed to be considered. Corrections were applied to the measured yield to account for two of these: dead time in the detectors and electronics, and randomly triggered Cherenkov events. Typical dead times were $\sim$30 ns. Applying these corrections resulted in an increase in the magnitude of the asymmetry for both targets. For hydrogen, there was a 2 ppm, or 10\%, increase, while for the deuterium the change was much larger at 12.6 ppm, or nearly 90\% of the uncorrected value. This large shift in the asymmetry for the deuterium data was due to the Cherenkov corrections, which were more significant because of high rates of $\pi^-$. Error analysis showed that for each data set the two corrections applied provided a negligible contribution to the uncertainty. 

The third rate-dependent effect, random CED$\cdot$FPD coincidences, was not included in the corrections applied. This effect was instead treated as an uncertainty. For each data set, an upper bound on the residual false asymmetry due to the presence of CED$\cdot$FPD randoms was estimated using information obtained from the elastic electron locus, where rate-correlated effects were more pronounced. The bounds determined from this analysis, 0.16 ppm for hydrogen and 1.2 ppm for deuterium, were assigned as uncertainties for the rate corrections.

% LINEAR REGRESSION
Corrections were also applied to account for any false asymmetry arising from helicity-correlated beam properties, including angle, position and energy of the electron beam. Because of the high quality of beam that was provided by the Jefferson Lab accelerator, helicity-correlated effects were negligible. As a result, the impact of these effects was also small ($A_{false}<$ 0.3 ppm). A conservative 100\% uncertainty was assigned for this correction.

%BEAM POLARIZATION 
The beam polarization was determined using a M\o ller polarimeter located upstream of the target \cite{Moller_NIMA}. Measurements taken periodically throughout the experimental run found the polarization to be steady at $P$ = 85.8 \% $\pm$ (0.07)$_{stat} \pm$ (1.4)$_{sys}$. Though the beam polarization was nominally longitudinal, there was a small transverse component present. However, the symmetrical nature of the $G^0$ spectrometer meant this component's impact on the asymmetry was negligible and no correction was necessary. Instead, using information determined through a measurement of the asymmetry with transverse beam and an estimate of detector misalignment, an upper bound on the false asymmetry was determined for each target and treated as an uncertainty. (See Table \ref{TabCorSum})

%BACKGROUNDS: INTRO
The most significant correction applied to the inelastic asymmetry was the background correction. Since the spectrometer was optimized for elastic, not inelastic, scattering, the yield in the inelastic region of the matrix contained a high percentage of background. In order to correct for backgrounds, a procedure was developed \cite{CapuanoThesis} that made use of both background measurements and simulated yields to determine the fractional contribution, or dilution factor ($f^{bg}_i$), of each process. The dilution factors were then used to subtract the background asymmetries ($A^{bg}_i$) from the cell average asymmetry ($A_{meas}$) according to
\begin{equation}
    A_{inel} = \frac{A_{meas} - \sum f^{bg}_iA^{bg}_i}{1 - \sum f^{bg}_i}~.
\label{DilACorDefn0}
\end{equation}

%BACKGROUNDS: PROCESSES
In any given CED$\cdot$FPD cell, there were up to five major processes contributing to the total yield and average asymmetry: electrons scattered elastically from the target liquid, electrons scattered inelastically from the target liquid, electrons scattered elastically or inelastically from the aluminum windows of the target cell, $\pi^0$ decay, and misidentified $\pi^-$. Note that the $\pi^-$ contribution here differs from that which is removed by the rate corrections, with the contamination resulting both from rate-independent effects causing false Cherenkov triggers and from $\delta$-ray production. The $\pi^0$ contribution to the background is due to electrons that are emitted through secondary processes following pion decays, primarily electron-positron pairs from decay photons interacting in the shielding.

\begin{figure}[ht]
  \centering
  \includegraphics[width=\columnwidth]{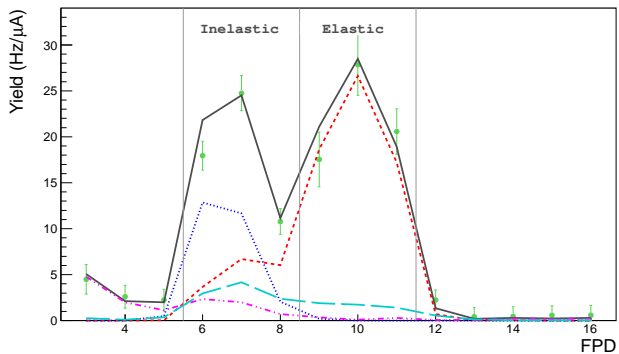}
  \caption{Contributions of the different processes present in CED 4, Octant 3, for the hydrogen data. The data are given as points and the solid curve represents the total yield as determined through background analysis. The individual processes are shown as curves: inelastic scattering (dot), elastic scattering (dash), $\pi^0$ decay (dash-dot), scattering from Al target windows (long dash). The vertical line is included to highlight cells located in the elastic and inelastic loci, as indicated by the labels. Note that the yields determined through background analysis are drawn as curves rather than individual points to make the plot more clear. The shape of the curves comes only from connecting the points together and has no other significance.}
  \label{BgCorFig}
\end{figure}

%BACKGROUNDS: DETERMINATION OF f_bg
The yields due to scattering from the target windows and from pion contamination were determined using data from special measurements made during the experimental run. For the three remaining processes, simulation was used to model the individual yield distributions across the detector acceptance while a fitting routine determined the appropriate normalizations. Before applying the fit, the target window and $\pi^-$ yields were subtracted from the total yield leading to a reduced yield, $Y_R$, consisting of elastic, inelastic and $\pi^0$ decay yields. $Y_R$ was then fit as a function of FPD for each CED, or across each row in the yield matrix shown in Fig. \ref{CoincMat}, by assigning scale factors to the simulated yields and allowing them to vary independently. Since our symmetrical spectrometer implies constant yields across the detector octants, all octants were fit simultaneously to find a single scale factor that represented an octant average for the CED. Once the scale factors were determined, a second fit of scale factor as a function of CED was performed for each process to ensure that the scale factors varied smoothly across the matrix. The yield contribution for a given process in a given cell was then defined as the simulated cell yield multiplied by the fitted scale factor for that cell. Details on the methods used to determine the cell-by-cell yield contribution from each process are given in \cite{CapuanoThesis}. Figure \ref{BgCorFig} shows the full background separation for a typical CED 
in a typical octant for the hydrogen data.

%BACKGROUNDS: ASYMMETRIES
Measured asymmetries determined from $G^0$ data were used for each of the background processes. An asymmetry of 0 $\pm$ 3 ppm, based on our measured $\pi^-$ asymmetry and experimental uncertainty, was used for both the $\pi^0$ decay and $\pi^-$ backgrounds. Our measured elastic electron asymmetry \cite{BckAngleLetter} was used with a scaling, determined through simulation, to account for the fact that the background in the inelastic locus is mainly from the elastic radiative tail, which has a lower average $Q^2$ than our elastic data. The scaling was computed cell-by-cell, with those closest to the elastic locus having the smallest scale factors. The locus average scaling was about 30\% for hydrogen and over 50\% for deuterium. For the empty target background, the dominant process was $\Delta$-electroproduction from aluminum. Although we were able to use data to determine the yields, the data did not have sufficient precision for an asymmetry measurement. Instead, the aluminum asymmetry was approximated using the measured deuterium asymmetry. This asymmetry differs slightly from the true aluminum asymmetry due to the differing proton to neutron ratio between the two nuclei, however, this difference is small compared to the error in the measurement. Note that it was not necessary to remove the empty target background from the deuterium data since the asymmetries were considered equal.

%BACKGROUNDS: SUMMARY OF RESULTS
 Table \ref{BgCorSum} summarizes the dilution factors and their corresponding background asymmetries. The total background fraction in the inelastic locus was (52.5 $\pm$ 4)\% for the hydrogen data and (64.9 $\pm$ 5)\% for the deuterium data. For both targets, the radiative tail of the elastic electrons was the dominant background, contributing 25\% of the hydrogen yield and 30\% of the deuterium yield. The higher total background fraction in deuterium was due both to the presence of $\pi^-$ contamination and to the widening of the elastic peak due to Fermi motion in the deuterium nucleus. Subtraction of the backgrounds results in a $-7.3$ ppm change in the hydrogen asymmetry and a $-12.6$ ppm change in the deuterium asymmetry. The background correction also results in an approximately 5 ppm systematic error due to the uncertainties from both the measured asymmetries and the dilution factors.

\begin{table}[h]
  \centering
  \begin{tabular}{l r r}
    \multicolumn{3}{c}{Hydrogen 687 MeV}\\
    \hline
    Process & $f_{bg}$ & $A_{bg}$\\
    \hline
    Elastic & 25.7 $\pm$ 0.4 \%& -14.5 $\pm$ 0.89 ppm\\ 
    Empty Target & 15.6 $\pm$ 0.8 \%& -43.6 $\pm$ 16 ppm\\
    $\pi^0$ Decay & 13.6 $\pm$ 3.6 \%& 0 $\pm$ 3 ppm\\
    $\pi^-$ & \emph{negligible} & 0 $\pm$ 3 ppm\\
    \hline
    \multicolumn{3}{c}{}\\
    \multicolumn{3}{c}{Deuterium 687 MeV}\\
    \hline
    Process & $f_{bg}$ & $A_{bg}$\\
    \hline
    Elastic & 31.0 $\pm$ 0.3 \%& -31.4 $\pm$ 2.2 ppm \\ 
    Empty Target & 9.0 $\pm$ 0.5 \%& --- \\
    $\pi^0$ Decay & 11.3 $\pm$ 3.2 \%& 0 $\pm$ 3 ppm\\
    $\pi^-$ & 11.1 $\pm$ 3.3\% & 0 $\pm$ 3 ppm\\
    \hline
  \end{tabular}
  \caption{Summary of the dilution factors and background asymmetries.}
  \label{BgCorSum}
\end{table}

%RADIATIVE CORRECTIONS & ACCEPTANCE AVERAGING
For the hydrogen asymmetry, two additional corrections were needed before comparing our results to theory. The first of these accounts for electromagnetic radiative effects by using simulation to compute corrections according to the procedure of Mo and Tsai \cite{MoTsai69} \cite{Tsai71}. The total correction was determined by comparing the simulated locus-average value of the asymmetry with and without radiative corrections included and was found to be (1.17 $\pm$ 0.6)\%. The uncertainty arises from the model chosen for the inelastic cross section. We then accounted for bias due to acceptance by comparing the locus average asymmetry, $\langle A(Q^2,W)\rangle$, to the asymmetry at the average kinematics, $A(\langle Q^2 \rangle, \langle W \rangle)$, and found this to be a (-1.6 $\pm$ 0.6)\% effect. The uncertainty here was determined by comparing the asymmetries computed with the two formalisms discussed in Section \ref{SecTheory}.

%FINAL SUMMARY OF CORRECTIONS
Table \ref{TabCorSum} summarizes the corrections applied, the systematic uncertainty assigned for each, and their impact on the asymmetry. Backgrounds constituted the largest correction to the asymmetry and the largest contribution to the uncertainty. Note that, in addition to its impact on systematic uncertainty, the background correction led to an increase in the statistical uncertainty as background events were removed. Additionally, since the hydrogen result made use of the deuterium result for the aluminum asymmetry, the systematic uncertainty and the large statistical error present in the deuterium asymmetry were factored in to the systematic uncertainty for hydrogen. The second largest impact was that of the rate corrections, with the larger contribution to the deuterium asymmetry due to generally higher scattering rates and greater contamination caused by high pion rates. The total uncertainty, including both systematic and statistical errors, is 7.4 ppm for the hydrogen asymmetry and 16 ppm for deuterium.

\begin{table}[h]
  \centering
  \begin{tabular}{l c r}
    \multicolumn{3}{c}{Hydrogen 687 MeV}\\
    \hline
    & dA (ppm) &$\sigma^{cor}_{sys}$ (ppm)\\
    \hline
    Rate Corrections        & -2.17 & 0.16 \\
    Helicity Cor. Beam      & -0.16 & 0.16 \\
    Beam Polarization       & -3.91 & 0.36 \\
    Transverse Polarization &   --- & 0.03 \\
    Backgrounds             & -7.33 & 4.93 \\
    EM Radiative Effects    & -0.39 & 0.20 \\
    Acceptance Averaging    & -0.55 & 0.20 \\    
    \hline
    Total $\sigma_{sys}$ &  & 5.11 \\
    Total $\sigma_{stat}$&  & 5.30 \\
    \hline
    \multicolumn{3}{c}{}\\
    \multicolumn{3}{c}{Deuterium 687 MeV}\\
    \hline
    & dA (ppm) & $\sigma^{cor}_{sys}$ (ppm) \\
    \hline
    Rate Corrections        & -12.6 & 1.20 \\
    Helicity Cor. Beam      & +0.25 & 0.25 \\
    Beam Polarization       & -4.66 & 0.43 \\
    Transverse Polarization & ---   & 0.02 \\  
    Backgrounds             & -12.5 & 5.52 \\       
    \hline
    Total $\sigma_{sys}$ & & 6.23 \\
    Total $\sigma_{stat}$& & 14.6\\
    \hline
  \end{tabular}
  \caption{Systematic uncertainty ($\sigma^{cor}_{sys}$) due to the corrections applied and each correction's impact (dA) on the average asymmetry. The total statistical uncertainty, $\sigma_{stat}$ is given for comparison purposes.}
  \label{TabCorSum}
\end{table}

\section{Results and Conclusion}

Upon applying the corrections outlined in the previous section, we arrive at the final measured asymmetries for the hydrogen and deuterium data. These asymmetries are
\begin{align}
  A_{inel} &= -33.4 \pm (5.3)_{stat} \pm (5.1)_{sys},\\
  A^D_{inel}&= -43.6 \pm (14.6)_{stat} \pm (6.2)_{sys}.
\end{align}
We found the deuterium asymmetry to be consistent with the hydrogen asymmetry within errors as would be expected. As discussed previously, due to the lack of a model for the deuterium asymmetry, no further information was extracted from $A^D_{inel}$. For the hydrogen asymmetry, however, theoretical input can be used to determine the form factor of interest in this measurement. 

The measured $G^A_{N\Delta}$ can be extracted from the asymmetry by first using the theoretical values of $A_1$ and $A_2$ discussed in Section \ref{SecTheory} to isolate the axial response, $A_3$. Subtracting off the vector portions of $A_{inel}$ leads to $A_3 = -0.69 \pm (5.3)_{stat}\pm (5.1)_{sys}\pm (0.7)_{th}$ ppm, where the subscript "th" denotes the theoretical uncertainty. The theoretical value for the resonant contribution calculated using the formalism of Musolf \emph{et al.} is $A_3=-$1.8 ppm, while the Matsui \emph{et al.} formalism leads to $A_3=-$1.7 ppm. The measured value for $A_3$ is consistent within errors with both of these values. While these results indicate that there are no major deficiencies in the theory, or in the assumption that the non-resonant contribution can be neglected, there is insufficient precision in the present measurement to select one prediction over the other. 

From this value of $A_3$, $G^A_{N\Delta}$ was found to be $-0.05\pm (0.35)_{stat}\pm(0.34)_{sys}\pm(0.06)_{th}$. Here the full theoretical uncertainty of 1.0 ppm has been included. The resulting $G^A_{N\Delta}$ is consistent, within errors, with the theoretical value of $-$0.196.

\begin{figure}[ht]
  \centering
  \includegraphics[width=\columnwidth]{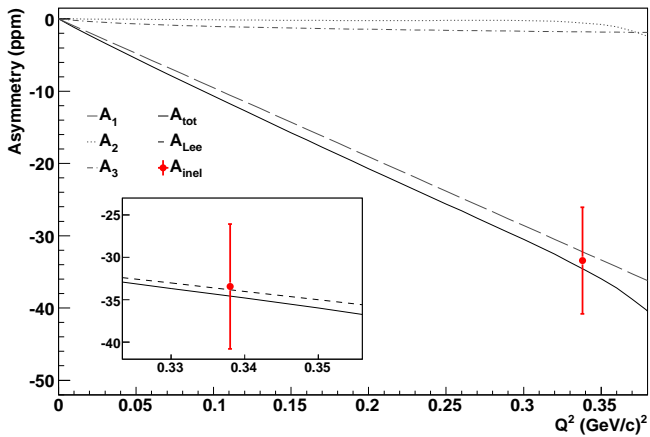}
  \caption{Measured hydrogen asymmetry (point) plotted with the total theoretical asymmetry, $A_{th}$, computed using the Musolf \emph{et al.} formalism (line). The asymmetry components $A_1$ (long dash), $A_2$ (dot) and $A_3$ (dash-dot) are also shown. \textbf{Inset:} Comparison of the measured asymmetry to the total $A_{th}$ computed using the formalisms of Musolf \emph{et al.} (line) and Matsui \emph{et al.} (dash).} 
  \label{AsyThWithData}
\end{figure}

Figure \ref{AsyThWithData} shows a comparison of the measured hydrogen asymmetry to the theoretical values computed using the definitions in Section \ref{SecTheory}. The statistical, systematic and theoretical uncertainties have been summed in quadrature to yield a single total uncertainty. The measured asymmetry, $A_{inel}$ is consistent with the total theoretical asymmetry, $A_{th}$ = $A_1 + A_2 + A_3$, within errors. 

In summary, we have measured the parity-violating asymmetry in inelastic electron-proton scattering and found excellent agreement with theoretical expectations based on the formalisms of Musolf \emph{et al.} \cite{PhysRept94} and Matsui \emph{et al.} \cite{TheoryLee}. In addition, the axial transition form factor, $G^A_{N\Delta}$, has been determined to be consistent with the theoretical formalism presented by Musolf \emph{et al.} \cite{PhysRept94} using the Adler parameterization. This represents the first neutral current measurement of the axial N$\rightarrow \Delta$ response from a nucleon target.

\section*{Acknowledgements}
We gratefully acknowledge the strong technical contributions to this experiment from many groups: Caltech, Illinois, LPSC-Grenoble, IPN-Orsay, TRIUMF, and particularly the Accelerator and Hall C groups at Jefferson Lab. CNRS (France), DOE (U.S.), NSERC (Canada), and NSF (U.S.) supported this work in part. We also thank Harry Lee and collaborators for providing their calculations at our kinematics.

%%\section*{References}

\bibliography{InelPaperBiblio}
\bibliographystyle{model1a-num-names}

\end{document}